\documentclass[twocolumn,trackchanges]{aastex631}

\usepackage{CJK}

\usepackage[version=4]{mhchem}
\usepackage{float,graphicx,multirow,tabularx}
\PassOptionsToPackage{dvipsnames,svgnames,table}{xcolor}
\usepackage[svgnames]{xcolor}
\usepackage{hyperref}
\usepackage{makecell}

\usepackage[utf8]{inputenc}
\usepackage[T1]{fontenc}
\usepackage{url}

\begin{document}
\begin{CJK*}{UTF8}{gbsn}


\title{Benchmarking Dimensionality Reduction Methods for High-Dimensional ALMA Image Cubes}

\author[0000-0002-8505-4459]{Haley N. Scolati}
\affiliation{Department of Chemistry, University of Virginia, Charlottesville, VA 22903, USA}
\affiliation{National Radio Astronomy Observatory, Charlottesville, VA 22903, USA}
\author[0000-0002-8932-1219]{Ryan A. Loomis}
\affiliation{National Radio Astronomy Observatory, Charlottesville, VA 22903, USA}
\author[0000-0001-9479-9287]{Anthony J. Remijan}
\affiliation{National Radio Astronomy Observatory, Charlottesville, VA 22903, USA}
\author[0000-0002-1903-9242]{Kin Long Kelvin Lee}
\affiliation{NVIDIA Corporation, 2788 San Tomas Expressway, Santa Clara 95051}
\altaffiliation{Current affiliation; worked done while at Intel Labs, 2111 NE 25th Ave, Hillsboro, OR 97124, USA}

\correspondingauthor{Haley N. Scolati}
\email{hns3nh@virginia.edu; hnscolati@chem.ubc.ca}


\begin{abstract}

High-dimensional astronomical data cubes provide a wealth of spectral and structural information that can be used to study astrophysical and chemical processes. The complexity and sheer size of these datasets pose significant challenges in their efficient analysis, visualization, and interpretation. In specific astronomical use cases, a number of dimensionality reduction techniques, including traditional linear (e.g. principal component analysis) and modern nonlinear methods (e.g. convolutional autoencoders) have been used to tackle this high-dimensional problem. In this study, we assess the strengths,  weaknesses, and nuances of various methods in their ability to capture and preserve astronomically-relevant features at lower dimensions. We provide recommendations to guide users in identifying and incorporating these treatments to their data, and provide insights into the computational scalability of these methods for observatory level data processing. This benchmark study uses publicly available archival ALMA data from a diverse sampling of source morphologies and observing setups to assess the performance and trade-offs between computational cost, image reconstruction accuracy, and scalability. Finally, we discuss the generalizability of these techniques in regard to data segmentation and labeling algorithms and how they can be exploited for advanced data product generation and streamlined archival analysis as we prepare to enter the era of the ALMA Wideband Sensitivity Upgrade.  

\end{abstract}

\keywords{Astronomy image processing(2306) --- Astronomy data analysis(1858) --- Astronomy data visualization(1968) ---  Interdisciplinary astronomy(804)}


\section{Introduction} \label{sec:intro}

Radio interferometric facilities, such as the Atacama Large Millimeter/submillimeter Array (ALMA) in the Atacama Desert of northern Chile, have made their data easily accessible through a comprehensive archive, lowering the threshold for scientific research at all levels \citep{Stoehr2014}. Over the past few decades, advances in instrumentation and imaging techniques have driven a rapid increase in high-dimensional data. With upcoming correlator upgrades and the addition of new wideband receivers, even greater challenges are expected in the processing and analysis of both spatial and spectral data. 

With the advent of large-scale observatory projects, such as the Next Generation Very Large Array (ngVLA) and the ALMA Wideband Sensitivity Upgrade (WSU), the expected dataload of the observatory will significantly increase. For context, the WSU alone will increase the bandwidth of ALMA to ${\sim}$1.2 million channels \citep{carpenter2022}. These projects and instrumentation upgrades create the need for efficient analysis, handling, and storage of astronomical data. 

For any set of scientific end goals, a standard data procedure is typically followed: the data undergoes preprocessing treatments, such as calibration, imaging, and classical preprocessing (e.g. normalization, standardization), the data is then dimensionally reduced, and finally the user can apply their technique of choice to achieve their end product. Although preprocessing pipelines have been extensively developed to incorporate automated tasks such as automasking, flagging, and generating imaging heuristics, similar progress has not been broadly made to incorporate dimensionality reduction within radio astronomy. 

In other areas of research, data processing procedures include a dimensionality reduction step where the data is reduced into lower-dimensional representations that are more manageable to process yet preserves the most important features, or information, of the original dataset. Some of these fields, such as remote sensing, have used statistical methods, including principal component analysis \citep[PCA;][]{Wold1987}, to analyze hyperspectral image data, whereas now these fields have improved to using state-of-the-art techniques, such as convolutional neural networks and vision transfomer-based architectures \citep{uddin2021, Adegun2023}. Astronomical studies have drawn from these advancements, experimenting with various machine learning (ML) and deep learning (DL) architectures for radio data tasks including object detection and classification \citep{DelliVeneri2023}, ALMA cube denoising and dimensionality reduction \citep{einig2023}, neural map-based clustering for astronomical structural properties \citep{Merenyi2017}, and predictive analysis of protoplanetary disks \citep{Telkamp_2022}. While each of these studies proved successful, each use case was highly specific to their respective application and required extensive method customization. It remains uncertain how generalizable ML and DL techniques are for ALMA data cubes overall, specifically whether these techniques can be applied more broadly to general ALMA uses.

In this work, we aim to gain insight into if common methods can scale appropriately to general use case ALMA data. Three linear statistical and one nonlinear DL method---principal component analysis (PCA), sparse PCA, non-negative matrix factorization (NMF), and a neural network autoencoder architecture---were sampled and evaluated on their compression efficacy while capturing and preserving essential astronomical features. We perform an initial ``screening'' with all three techniques, benchmarking their performance with datasets from two well-studied source morphologies: star forming regions and protoplanetary disks. Subsequently, we apply the NMF and autoencoder methods to two additional morphologies to assess their effectiveness with distinct image features and dataset properties including low- and high-spectral resolution. Through reconstruction analysis, our goal is to understand which types of low-dimensional latent spaces provide the most informative and interesting representations. 

This article is organized as follows. Section 2 provides a description of ALMA data cubes, including format and generation process. General overviews of each source morphology with the structural characteristics needed to be captured in a “successful” image reconstruction are also provided. Section 3 highlights the dimensionality reduction methods used with general overviews of each method's theory and explanations as to why we chose to probe that method. In Section 4, we present the dimensionality reduction results using select astronomically-relevant metrics. We provide insights for the generalizability of these methods for specific science use cases in Section 5 and conclude our findings in Section 6 with a discussion of algorithm complexity, and scalability, as well as look forward to future considerations as we enter into the era of the ALMA Wideband Sensitivity Upgrade (WSU).


\section{Data set description} \label{sec:data}

\begin{figure*}
\centering
\includegraphics[width=0.8\textwidth]{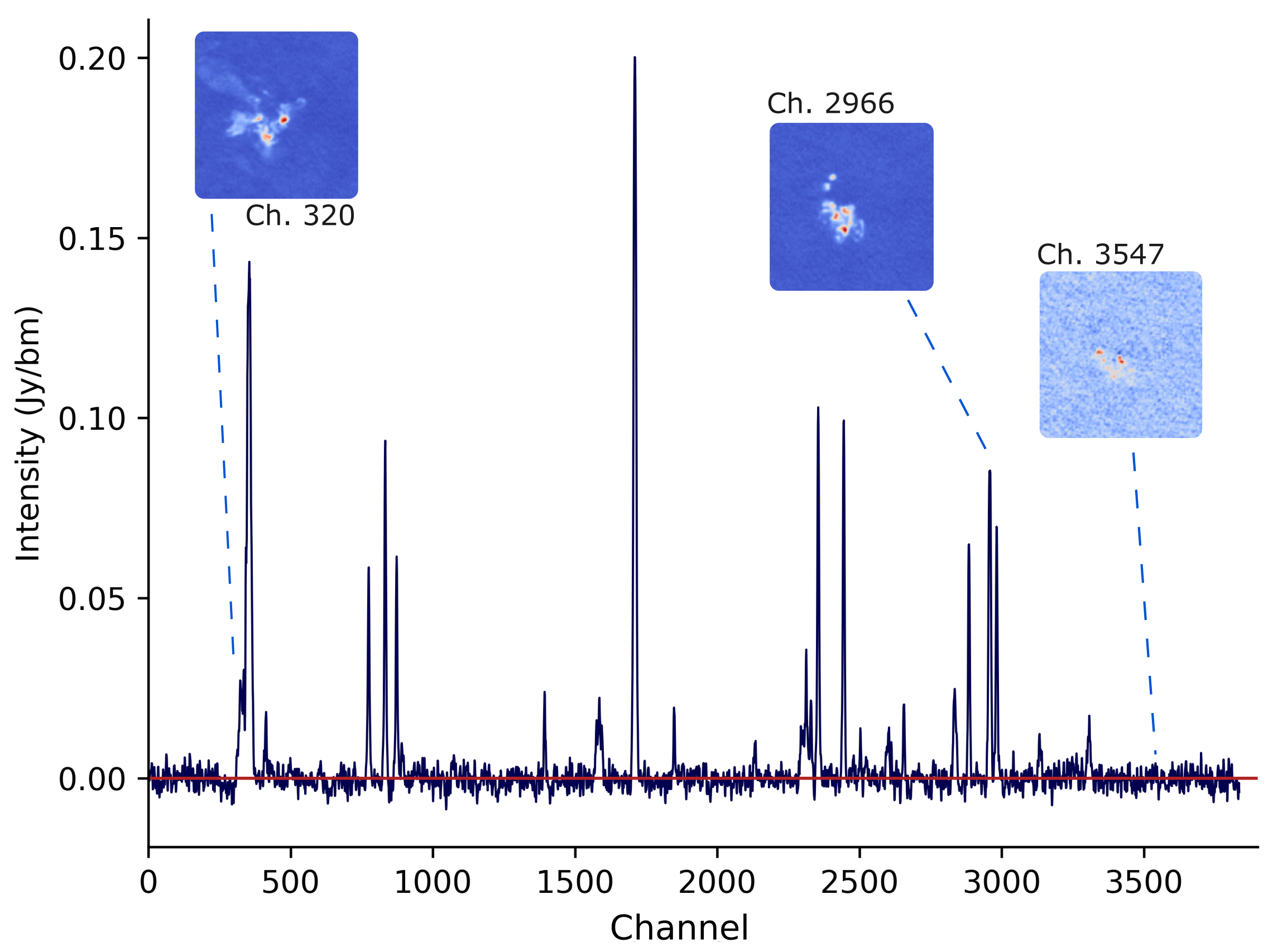}
\caption{G34.30 spectrum depicting the channels used for image reconstructions. The moment 0 maps shown are the corresponding emission along the channel (offset frequency) axis.}
\label{fig:spectrum}
\end{figure*}

Interferometric image cubes are generated by deconvolving the inverted Fourier transform of the measured visibilities using the \textsc{clean} routine, originally presented in \citet{1974AAS...15..417H}. The resulting three-dimensional image cube consists of voxels---the three-dimensional analog of a two-dimensional pixel---that are often projected into lower dimensional representations. Slicing along the frequency axis, the cube is made up of a series of images (channels) over a frequency range that depict the emission intensity (Figure \ref{fig:spectrum}). These images can be integrated into moment maps to give detailed spatial and kinematic information. When extracted and collapsed along the spatial axis, the pixels yield molecular spectra. If we compare ALMA image cubes to a typical dataset used in computer vision—the field of interpreting image and video data with machine learning—ALMA data cubes are highly dimensional. Computer vision datasets segment images into a grid of pixels where each pixel is a numerical value representing that gridpoint's color using its RGB code. Alternatively, ALMA data cubes contain a third dimension that enables spectral information to be stored in both frequency and intensity domains. As a result, additional considerations need to be made when reshaping and collapsing the cubes along their axes to ensure critical information is not lost.

\begin{deluxetable*}{llcccccc}
\label{table:datasets}
\tablecaption{Characteristics and observing setups of sampled source morphologies.}
\tablecolumns{8}
\tablewidth{\columnwidth}
\tablehead{                                                                            
     \colhead{\shortstack[c]{Source}} &
     \colhead{\shortstack[c]{Source type}} &
     \colhead{\shortstack[c]{\rule{0pt}{10pt}Spectral \\ complexity}} &
     \colhead{\shortstack{$T_{ex}$ \\ (K)}} &
     \colhead{\shortstack{Line Width \\ km $s^{-1}$}}  &
     \colhead{\shortstack[c]{Array configuration}} 
          }
\startdata
G34.30+0.20  & Hot core  & dense   & $\sim$100-300~K  & $\sim$6  &  12m  &  \\
LkCa 15      & Disk      & sparse  & $\sim$10-80~K  & $\sim$10  & 12m  &  \\
IRC+10216    & AGB star  & sparse  & $\sim$10-40~K  & $\sim$25  &  12m &  \\
B33          & PDR       & dense   & $\sim$15-100~K  & $\sim$2 0.22  &  12m + 7m + TP  &  \\
\enddata
\end{deluxetable*}

\begin{figure*}
\centering
\includegraphics[width=0.8\textwidth]{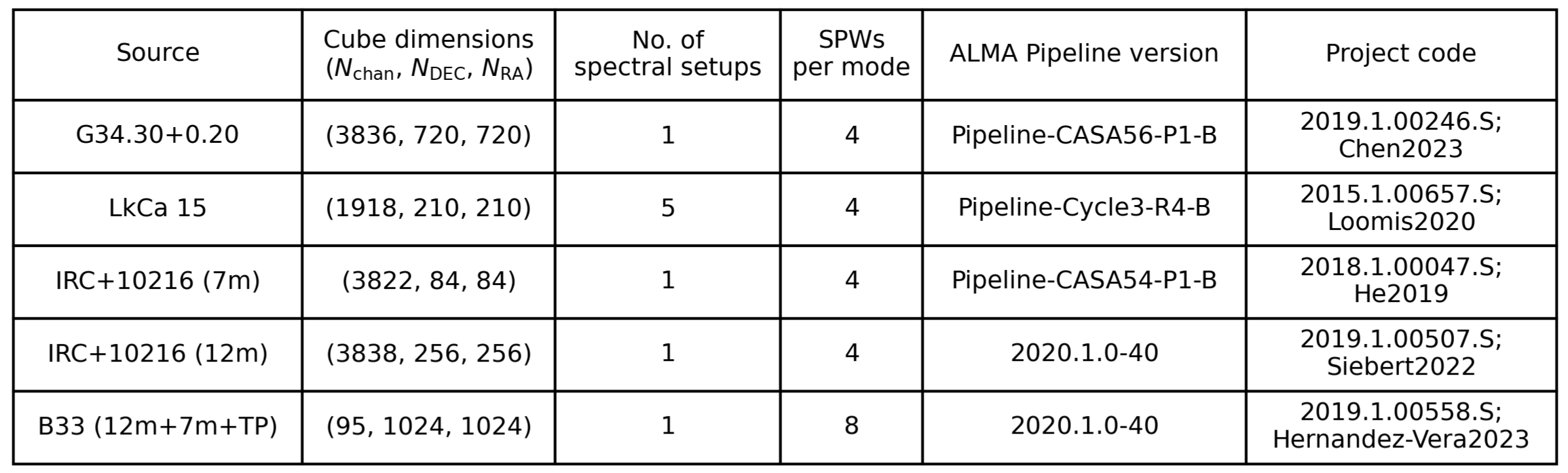}
\caption{Characteristics and observing setups of sapled source morphologies. \small \textsuperscript{\textdagger}ALMA pipeline and branch version can be referenced at \href{https://almascience.nrao.edu/processing/science-pipeline}{https://almascience.nrao.edu/processing/science-pipeline.}}
\label{fig:cube_data}
\end{figure*}

\subsection{Sampled astronomical sources and observing setups\label{subsec:sources}}
To assess the effectiveness of compression methodologies, we selected a variety of sources based on morphology and data set types. The chosen sources represent common morphologies within four subfields---star-forming regions, protoplanetary disks, asymptotic giant branch stars, and extended photodissociation regions. This selection was intended to provide a diverse benchmark of astronomical data, encompassing a range of image cube sizes, spectral complexity, and observing setups. Each subsection below offers a brief description of the source, including physical characteristics, the archival dataset used, and the key features deemed essential for retention during dimensionality reduction. A summary of the physical parameters for each source are provided in Table \ref{table:datasets} and Table \ref{fig:cube_data} details the image cube properties for each dataset. Compression metrics will be discussed in Section 4.

\subsubsection{High mass star forming region: G34.30+0.20\label{subsec:G34}}

Hot molecular cores are part of active star-forming regions characterized by their small and dense structure of warm molecular gas \citep{DeSimone2020}. This makes them particularly molecular rich sources yielding very line dense millimeter and submillimeter spectra \citep{Nazari2023, Rivilla2017}. Despite the potential of cataloging extensive chemical inventories, line blending and spectral overlap hinders efficient parsing and analysis for possible new detections. We chose the star-forming region G34.30+0.20 as our spectral-rich benchmark using archival data from the CoCCoA hot core survey \citep{Chen2023}.     

The data cubes are characterized by variable molecular emissions ranging from intensely bright, concentrated knots to very wispy, diffuse regions extending away from the cores. The spectra are line confusion limited, with lower \textit{J}-level rotational transitions and blended molecular lines contributing to a dense spectral line forest. Self-absorption and baseline over-subtraction appear to be present in the cubes, given the sharply decreasing and negative spectral values surrounding the brighter regions of emission. Optically thick species present additional challenges where saturated intensities cause non-linear effects. These qualities coupled with linear and non-linear processes highlight some of the intricacies that are crucial to be captured in the image reconstructions.

\subsubsection{Protoplantary disk: LkCa 15\label{subsec:disks}}

LkCa 15 is a young T Tauri star located in the Taurus Molecular Cloud \citep{Kraus2012, Brown2018}. It is surrounded by a protoplanetary disk, making it an ideal source for studies that aim to disentangling disk structure and astrophysical processes, as well as volatile disk chemistry and molecular composition \citep[e.g.][]{Loomis2020, Kastner2018}. The gas within disks is quite cold, resulting in the molecular species within the midplane of the disk to be locked-up on icy grains and causing their rotational transitions to ``freeze out'' and not be observable within the microwave regime \citep{Cuppen2017}. This makes targeted line surveys with radio astronomy particularly difficult and chemical inventories incomplete \citep{Loomis2020}. Even with incredibly sensitive data from ALMA or single dish telescopes, additional tricks such as matched filtering \citep{Loomis2018} are needed to tease out weaker signals consequence from the inherent low column densities.  

The resulting data cubes are line sparse, with sporadic emission from small volatile molecules such as \ce{N2H+} and \ce{C2H}. Here we opted to use the LkCa 15 archival data from \citet{Loomis2020} as our training data. The reconstructions of this dataset would need to capture the wide range of pixel values that correspond to bright emission that is only found within a few channels across each spectral window. The data set consists of a total of 20 spectral windows using 5 correlator spectral setups covering ~42 GHz of ALMA Band 7 with gaps in the 306--310 GHz and 318--322 GHz frequency ranges. The data was initially calibrated by ALMA/NAASC staff with later self-calibration and imaging performed by \citet{Loomis2020} using the CASA 4.3.1 software and \textsc{clean} algorithm.

\subsubsection{Asymptotic giant branch star: IRC+10216\label{subsec:AGB}}
Asymptotic giant branch (AGB) stars represent the final phase of stellar evolution for luminous evolved cool stars (1--8 $M_\odot$) prior to forming into planetary nebulae. These sources are characterized by their bright inner cores and expansive circumstellar envelopes, formed through significant mass loss \citep{Sahai2011}. AGB stars are observationally recognizable by their dusty envelopes extending outward into circular, ripple-like patterns. Several studies have indicated binary companions have profound influence on the mass loss and stellar winds of AGB stars, affecting the rich photo-induced chemistry housed within these envelopes \citep{Siebert2022, Cernicharo2015, Guelin2018}. Mapped emission depicts bright arcs with scattered clumps that outline the regions of the extended envelopes. 

For this study, IRC+10216 (also known as CW Leo) was chosen as it is one of the most well-studied AGB stars \citep[e.g.][]{He2019}. A 12-m dataset was primarily used in the dimensionality reduction assessment, however reconstructions were also generated using a 7-m dataset as the ALMA 7-m configuration is more compact than the 12-m configuration, causing different spatial effects, resolution, and artifacts in the images due to variation in antenna spacing.

\subsubsection{Photodissociation region: Bernard 33\label{subsec:pdr}}

Bernard 33 (B33, or more commonly known as the Horsehead Nebula) is a well-studied photodissociation region (PDR) given its brightness in the mid-infrared. B33 is broken into three distinct regions that differ significantly in their physical conditions (i.e. gas temperature, density, etc.). These conditions vary as a function of the visual extinction, indicating drastically changing chemistry from one region to the next and making it an excellent PDR probe for astrochemical and physical studies \citep{Le_Gal_2017}. The source is chemically rich with line surveys, such as the WHISPER survey (PI: J. Pety), and the encouraged additional targeting  of complex organic and prebiotic molecules \citep{Guzman2013, Guzm_n_2014}.          

The dataset used is mapped emission of the CO \textit{J} = 3--2 (345.796 GHz) transition reported in \citet{Hernandez-Vera2023} and is the only dataset that uses Group Observing Unit Set (GOUS)\footnote{Observing Unit Set with one or more Member Observing Set Unit(s) are combined during data processing.} imaged data. Observations were taken with the 12m array and ACA in different execution blocks and merged into a single measurement set for imaging. Visibilities were calibrated by the ALMA Observatory using the standard ALMA Pipeline v2020.1.0.40 \citep{Hunter2023}. Imaging was performed by \citet{Hernandez-Vera2023} using the \texttt{tclean} function with the CASA package. The CO emission was observed at the edge of the PDR, causing the emission to span the entire field of view of the image. As a result, these channels maps differ from the other datasets in that the majority of pixel space consists of emission rather than noise.


\section{Sampled methods \& experimental design} \label{sec:meth}

Four data encoding methods were explored for dimensionality reduction, and their effectiveness was assessed by examining the fidelity of reconstructed channel maps. A short description of each treatment and respective implementation are provided here.

\subsection{Experimental design \label{subsec:expDesign}}

In order to provide a thorough yet targeted benchmark of the selected treatments, we adopted the following experimental approach. All methods were first evaluated using the G34.30 star-forming region dataset and the LkCa 15 protoplanetary disk dataset. These datasets were chosen based on their well-studied, yet physically contrastive morphologies: G34.30 being spectral-line rich and a line forest of blended emission structures, while LkCa15 contains very few spectral lines, yet has very well-behaved emission. Upon analysis, two additional sources---B33 and IRC+10216---were then tested with the most successful methods. This approach enables a comprehensive analysis of each compression method, while also allowing exploration of supplementary datasets and use cases.    

\subsection{Methods\label{subsec:methods}}

\noindent {\bf Principal Component Analysis:} Principal components analysis (PCA) is a statistical approach of finding correlations among features in high-dimensional data via decomposition into a linear combination of orthogonal basis vectors (components). When truncated, lower dimensional vector spaces capture global structure by compromising reproduction of real variations and noise in the data. PCA is one of the most popular conventional methods of data compression as reducing the data to a lower-dimensional space can yield easier visualization and understanding with minimal information loss. PCA inherently orders the components by importance, allowing users to easily determine the number of components needed to capture the most important features of the data; however, interpretability is not always possible if the features are independent from one another. The extracted components are linear combinations of the input features, leading to dense eigensolutions with non-zero coefficients. Although the majority of the data can be reconstructed using a few, ranked components, the maximum variance captured within those components typically renders them difficult to interpret. 

PCA is notably used in denoising algorithms for medical imaging \citep{Does2019} and for rank-based analysis in high-dimensional biological data visualization \citep{Moon2019}. PCA was first used to denoise spectral lines in stellar data by \citet{Carroll2007}, and has since been used as a mulitvariant technique to decompose astronomical data for various investigations including molecular cloud structure \citep{Gratier2017}. The implementation of PCA used in our study corresponds to that of \textsc{scikit-learn} (sklearn) \citep{Pedregosa2011}. \linebreak


\noindent {\bf Sparse PCA:} Sparse PCA is a matrix factorization variation of PCA that introduces sparsity into the regression algorithm with the goal to produce more interpretable components \citep{Zou2006}. Sparsification encourages more of the linear combination coefficients to be zero, which in contrast to PCA, in theory leads to improved interpretability as fewer components are used and can be subsequently re-mapped to real features.

The degree of sparsity, $\alpha$, is controlled by introducing a penalty term where larger values of $\alpha$ result in sparser components (more zero-valued coefficients). How sparsity is applied is dependent on the data, model structure, and desired end goal analysis. However, despite the interpretability gained, the resulting principal components are not guaranteed to be orthogonal, posing potential issues such as non-unique solutions and complications to the iterative optimization routine leading to the algorithm converging to solution within a local minimum, or not converging at all \citep{Camacho2020, Guerra-Urzola2021}. 
\linebreak


\noindent {\bf Non-negative Matrix Factorization:} Non-negative matrix factorization (NMF) is a linear dimensionality reduction method that decomposes a multivariate matrix of only non-zero elements into two lower dimensional matrices. The product of the lower rank matrices are used as an approximation of the original, higher dimension matrix \citep{NIPS2000_f9d11525, Hien2015}. For image data, the input image ($\mathbf{A}$) is decomposed into a feature matrix ($\mathbf{W}$) and a matrix of coefficients ($\mathbf{H}$) that stores the weights of $\mathbf{W}$, projecting an approximation of the original image to this lower dimensionality. Due to its non-negativity constraint, the components extracted by NMF tend to be more interpretable compared to other methods like PCA.   

NMF is commonly used for dimensionality reduction in preprocessing and has been coupled with other methods in astronomical studies including the extraction of "pure" spectral components in hyperspectral data cubes \citep{Boulais2021} and as a post-processing technique to identify disk morphologies in image data \citep{bin2018}. Notably, the latter demonstrated that NMF surpassed current modeling methods and overcame limitations in post-processing imaging methods for extracting emission signal used to detect circumstellar disks. This suggests NMF to be a viable dimensionality reduction method to sample using general ALMA image datasets. \linebreak

\begin{figure*}
\centering
\includegraphics[width=0.8\textwidth]{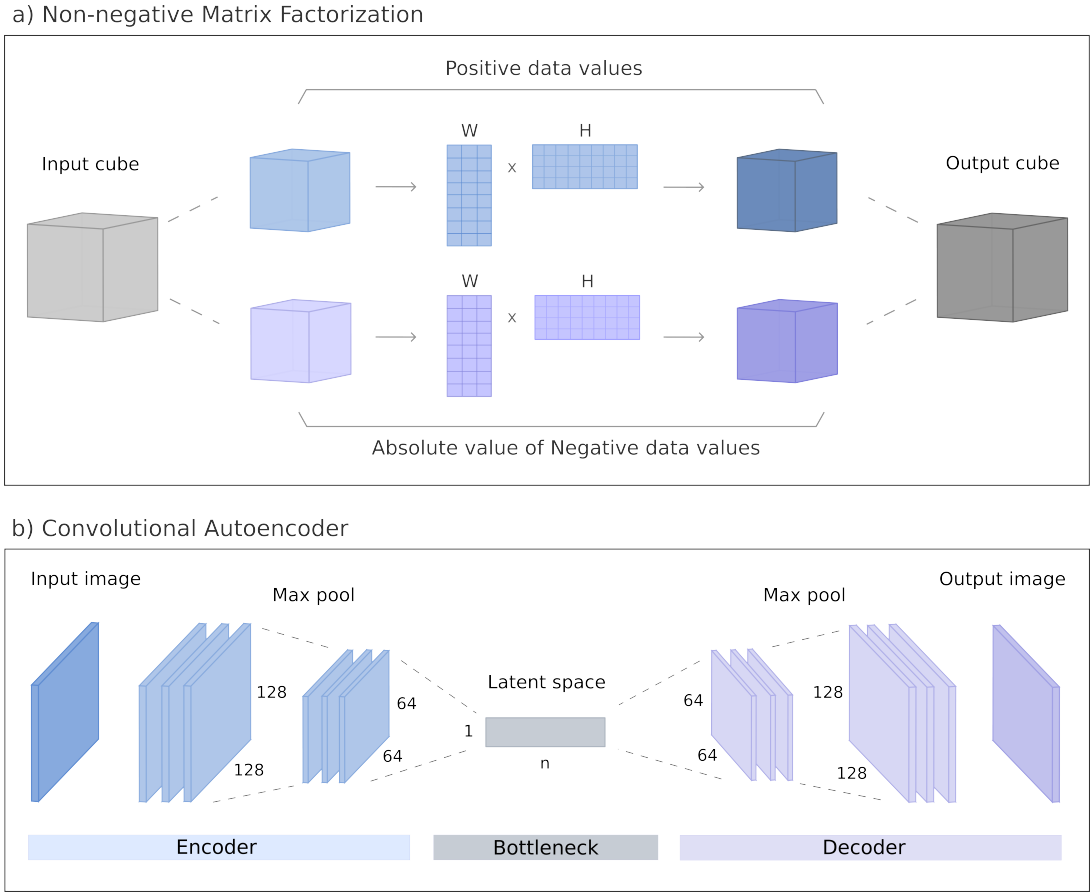}
\caption{Block diagrams of the (a) non-negative matrix factorization method and (b) symmetric convolutional autoencoder constructed for this work. (a) The input image cube was separated into a positive cube and a negative cube, where the absolute values within the negative cube were used to perform NMF. The resulting cubes (where the absolute values were returned to their negative values) were recombined to give the final reconstructed image cube. (b) The autoencoder architecture uses eight 15x15 pixel kernels to produce the convoluted images from a latent space of size 1x\textit{n}, where \textit{n} is the number of components. Max pooling is performed prior to each downsampling step, where the image dimensions in pixels in depicted above.}
\label{fig:block_v3}
\end{figure*}

\noindent {\bf Convolutional Autoencoders:} Neural networks are powerful tools for self-supervised feature extraction on large datasets. A neural network architecture corresponds to a directed acyclic graph that describes a sequence of transformations on input data. The sequence comprises ``layers'' that describe a set of freely composed functions applied to the data, ranging from matrix multiplication with learnable weights, regularization techniques, and non-linear element-wise transformations \citep{Ballard1987, LeCun2015}.

In this work, we use a convolutional autoencoder architecture, which specializes in treating image data owing to its highly parameter efficient structure \citep{Wilamowski2009}. Each layer within this architecture comprises a convolution kernel, and pooling operation, and a non-linearity: concretely, this corresponds to cross-correlation of an input image with a set of learnable 2D weight matrices which produces a ``feature map'' that is subsequently downsampled and gated. Figure \ref{fig:block_v3} is a block diagram that represents how these convolution layers are composed to produce a symmetric autoencoder: the encoder progressively transforms and downsamples the image into a lower dimensional latent space, while the decoder uses the latent encoding/embedding to reconstruct the image. The architecture is trained to minimize the cross-entropy loss, and through backpropagation, the encoder and decoder work cooperatively to reproduce training images accurately---the decoder needs feature rich or ``useful'' embeddings from the encoder in order to produce images of sufficient fidelity.

The autoencoder architecture used in this study is simple in design compared to other specialized and customized architectures \citep[e.g.][]{Schmidt2022,DelliVeneri2023}. In these examples, convolutional autoencoders were used for specific astronomical use cases whereas our goal is to asses these methods in their ``off-the-shelf'' forms to gain a wholistic understanding of their general utility within the field.

\section{Dimensionality Reduction Results} \label{sec:results}

Here we present the dimensionality reduction results using the four previously described compression methods. Table \ref{table:dim_table} provides the initial, downsampled, and latent cube dimensionalities for the sampled source datasets. All datasets were obtained using the ALMA archive. The FITS files were imported into a script to undergo initial downsampling and removal of NaN values. The final reconstructed products for PCA, Sparse PCA, and NMF were saved in NumPy binary format (\textit{.npy}), while the autoencoder outputs were stored as PyTorch model files (\textit{.pth}). Initial screening reconstruction generation of PCA, Sparse PCA, and NMF were performed using the Google Colab node equipped with an Intel Xeon CPU (2 vCPUs) and 12GB of RAM while final algorithm application and reconstruction generation presented in this work were performed using a local workstation equipped with an AMD Radeon RX 6500 XT GPU and AMD Ryzen 7 5800X 8-core processor. All autoencoder computations were performed using a local workstation at the National Radio Astronomy Observatory equipped with an NVIDIA GeForce RTX 2080 Ti GPU running Red Hat Enterprise Linux 8 (RHEL 8).

\begin{figure*}
\centering
\includegraphics[width=0.8\textwidth]{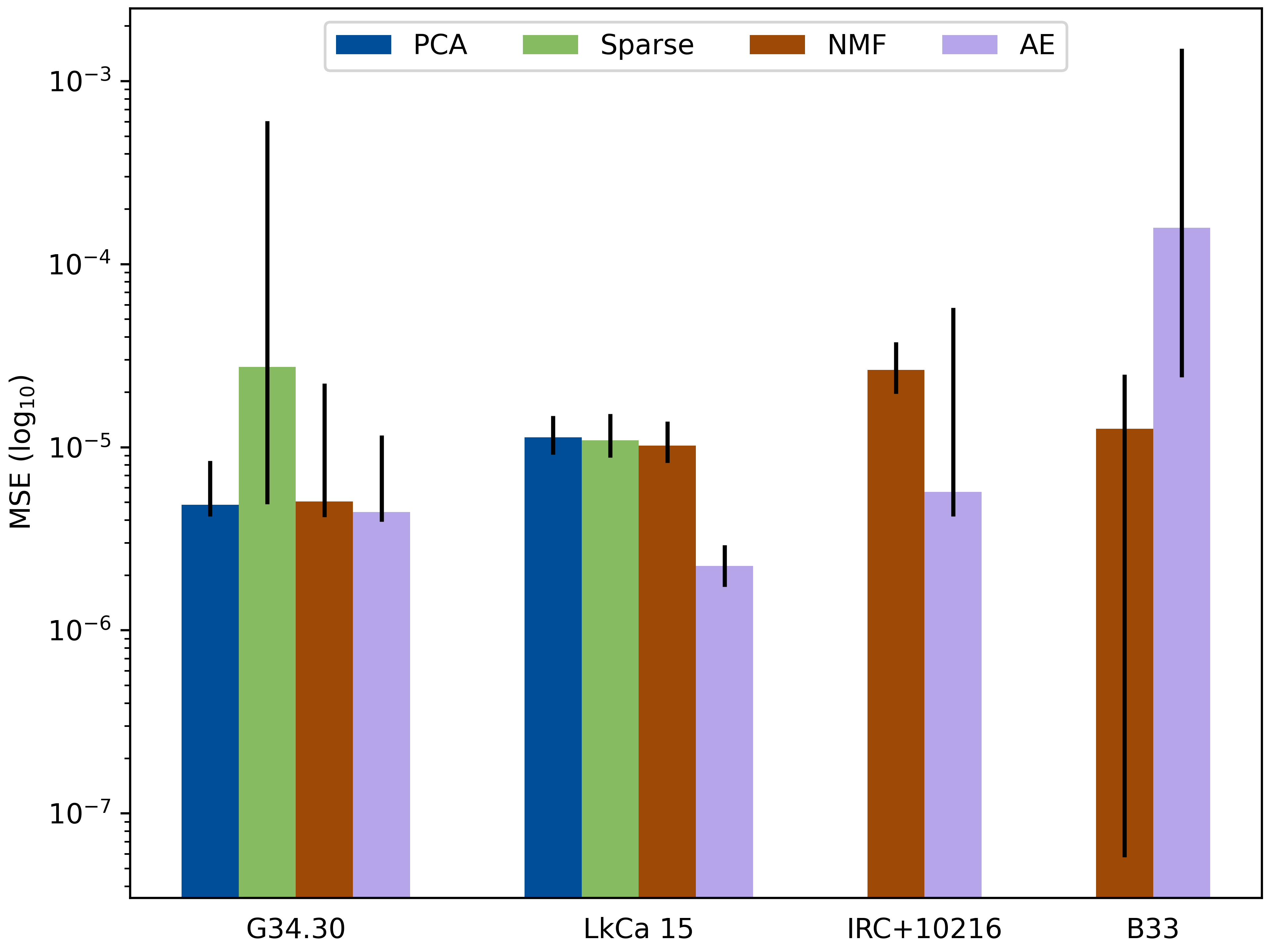}
\caption{Calculated mean squared error of reconstructed images with respect to the methods sampled for each source. The MSE was calculated on a per-channel basis. Error bars indicate the upper and lower bounds of the error for each method.}
\label{fig:mse_bar}
\end{figure*}

PCA and Sparse PCA were applied to the G34.30 and LkCa 15 datasets while NMF and the autoencoder were applied to all four datasets as they were deemed the two most ``successful'' compression treatments, based on the G34.30 and LkCa 15 results. Sparse PCA yielded unremarkable results and will not be discussed in detail here (see Appendix \ref{sec:AppB}). Both qualitative and quantitative metrics were used to assess the effectiveness of the methods. Reconstructed channel maps visualize image quality post compression. Residual and residual error of the reconstructions provide metrics of reconstructed image fidelity, indicating regions of greatest deviation from the ground truth images. The mean squared errors (MSE) were calculated for each dataset and provided in Figure \ref{fig:mse_bar}.

\begin{table*}[ht]
\centering
\caption{Initial, downsampled, and reduced ALMA image cube dimensions ($N_{\mathrm{chan}}$, $N_{\mathrm{DEC}}$, $N_{\mathrm{RA}}$) for each dataset.}
\label{table:dim_table}
\small
\begin{tabular}{lcccc}
\hline
 & \multicolumn{4}{c}{Dimensionality} \\
\cline{2-5}
Source & Initial & Downsampled & Latent NMF W;H & Latent AE \\
\hline
G34.30+0.20      & (3836, 720, 720) & (3836, 360, 360)\textsuperscript{\dag} & (3836, 50); (50, 32400) & (1, 200) \\
LkCa 15          & (1918, 210, 210) & (1918, 100, 100) & (1918, 50); (50, 10000) & (1, 200) \\
IRC+10216 (12m)  & (3838, 256, 256) & (3838, 150, 150)\textsuperscript{\dag} & (3838, 50); (50, 22500) & (1, 200) \\
B33 (12m+7m+TP)  & (95, 1024, 1024) & (95, 500, 500)\textsuperscript{\dag} & (95, 50); (50, 62500) & (1, 50) \\
\hline
\multicolumn{5}{c}{No. of Independent Samples} \\
\hline
Source & Initial & Downsampled & Latent NMF & Latent AE \\
\hline
G34.30+0.20      & $1.99\times 10^9$ & $1.38\times 10^6$ & $1.81\times 10^6$ & 200 \\
LkCa 15          & $8.56\times 10^7$ & $1.92\times 10^7$ & $5.96\times 10^5$ & 200 \\
IRC+10216 (12m)  & $2.52\times 10^8$ & $8.64\times 10^7$ & $1.32\times 10^6$ & 200 \\
B33 (12m+7m+TP)  & $9.96\times 10^7$ & $2.38\times 10^7$ & $3.13\times 10^6$ & 50 \\
\hline
\end{tabular}

\footnotesize
\textsuperscript{\dag}Denotes additional striding over each channel where only every other pixel in each dimension is sampled.
\end{table*}


\subsection{PCA Reconstructions of G34.30 and LkCa 15}
PCA was applied to the G34.30 and LkCa 15 datasets. The initial PCA approach encoded information from the entire cube, including both the true sky emission as well as thermal and non-thermal noise backgrounds, leading to these features all being present in the images reconstructed from the lower dimensional latent space. This was most pronounced in channels with lower SNR where emission levels were closer to the noise threshold, resulting in more aggressive reconstructed noise interference. To mitigate this, a mask was applied to the data cube to differentiate the molecular emission from the noise. A gaussian filter was used to blur the edges of the initial mask to incorporate lower laying signal and weak, extended emission around the mask.

\begin{figure*}
\centering
\includegraphics[width=0.8\textwidth]{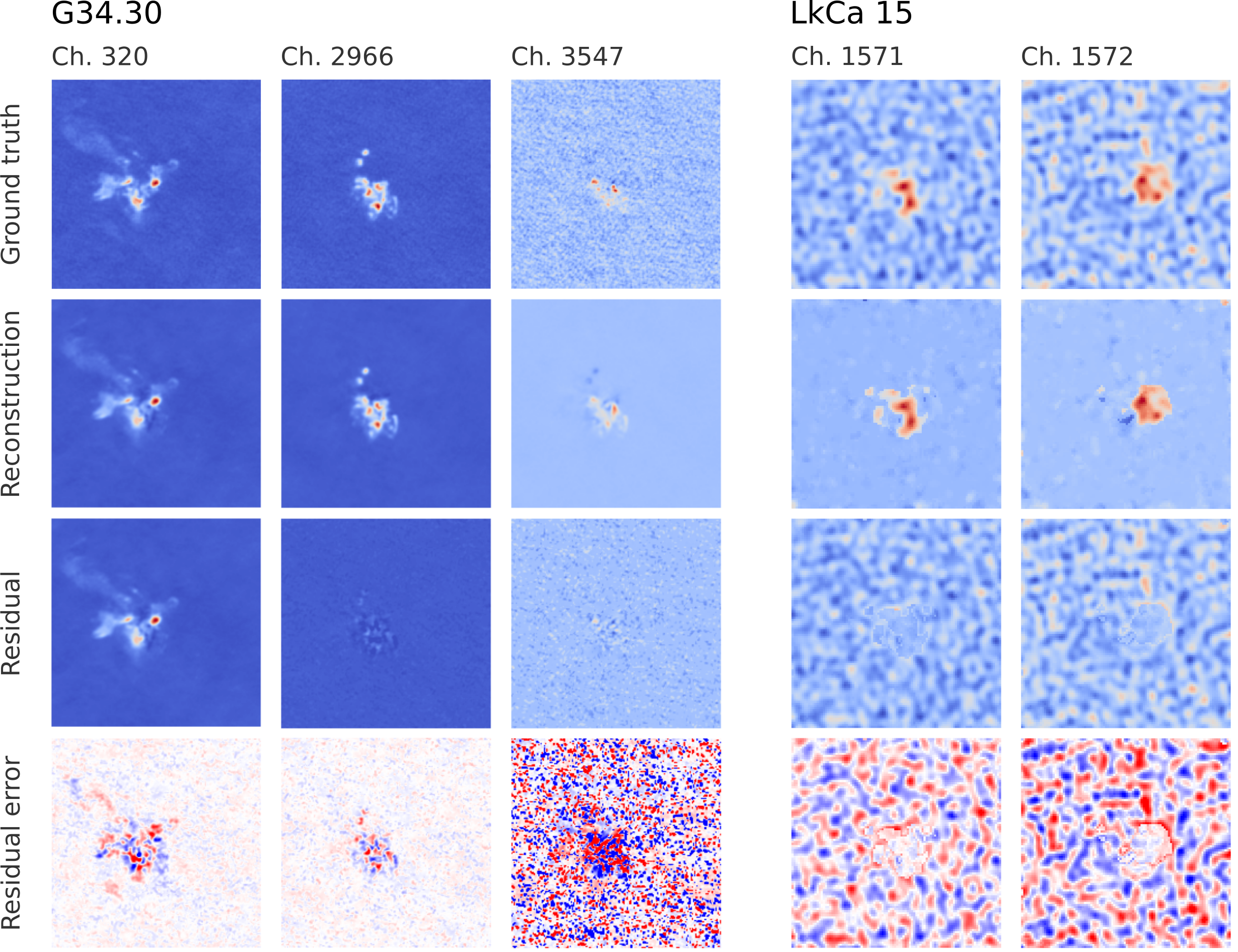}
\caption{PCA reconstructions, residuals, and residual errors with respect to ground truth images of select channels from the G34.30 (a) and LkCa 15 (b) datasets. The residual images were generated by taking the difference of the ground truth image and corresponding reconstruction. The color scale for the residual error panels in channels 320 and 2966 of G34.30 are 10\% of the maximum signal intensity, while the residual error for channel 3547 of G34.30 and channels 1571 and 1572 of LkCa 15 are 40\% of the maximum signal intensity.}
\label{fig:pca_residuals}
\end{figure*}

The resulting reconstructed channel maps and residuals demonstrated promising results that PCA can effectively reduce image dimensionality of ALMA data cubes with comparable errors to the other sampled methods. A total of 50 components were used in the reconstructions for both datasets where computations were performed using the local AMD workstation with typical execution times on the order of minutes. In the G34.30 dataset, PCA accurately captured emission structure and intensities in high signal channels with minimal error. After applying the mask, PCA sufficiently reduced thermal noise in low SNR channels; however, the reconstruction of the general emission structure was not consistent. These channel maps were often reimaged with ``ghost'' artifacts of stronger emission morphologies (i.e. channels 2966 and 3547 in Figure \ref{fig:pca_residuals}). 

Similar ``ghosting'' effects were anticipated in the LkCa 15 dataset as it is lower in signal resolution compared to the G34.30 dataset. Areas of lower-intensity emission were not effectively captured by the blurred mask due to the pixels being close in magnitude to the noise floor; although, the overall shape of the emission was reconstructed with fewer artifacts. The sparsity of lines in LkCa 15 dataset and limited number of channels containing emission suggests potential overfitting of the model. Figure \ref{fig:pca_residuals} provides the PCA-reconstructed channels, residuals, and residual errors for the selected channels in both datasets. 

Our PCA results demonstrated its utility in generating lower-dimensional images and proved that standard dimensionality reduction methods can be applied to astronomical datasets. It was most effective with well-resolved, signal-dominated spectral cubes. Performance was less successful with spectral-sparse cubes and channels with low signal-to-noise ratios.  Future improvements may consider training across multiple spectral windows to enhance sampling and reduce overfitting. These observations, as well as considering the required initial manual masking and significant computational demands, highlight the need for careful consideration of the applicability of PCA across diverse datasets and suggest the use of PCA is appropriate only for specific types of datasets and use cases.

\subsection{Sparse PCA Reconstructions of G34.30 and LkCa 15}

A manual grid search evaluated a small range of $\alpha$ values and number of components to identify the optimal settings for reconstructing lower-dimension channel maps. Based on the results from reconstructions, residual analysis, and computational time, we selected $n=10$ and $\alpha=0.01$ for the G34.30 dataset and $n=50$ and $\alpha=0.01$ for the LkCa 15 dataset (Appendix \ref{sec:AppA}). Sparse PCA demonstrated general limitations in channels with low signal-to-noise and datasets with less emission. It failed to readily capture low-resolution spectral features (particularly extended features just above the noise threshold) and accurate emission intensities. This was reflected in the channel reconstructions as well as the large MSE for the G34.30 dataset (Figure \ref{fig:mse_bar}). It was determined that Sparse PCA was not well suited for astronomical data cubes without significant hyperparameter optimization. Therefore, the method will not be further explored, however preliminary results and minor discussion can be found in Appendix \ref{sec:AppA}.

\subsection{NMF Reconstructions}

\begin{figure*}
\centering
\includegraphics[width=0.8\textwidth]{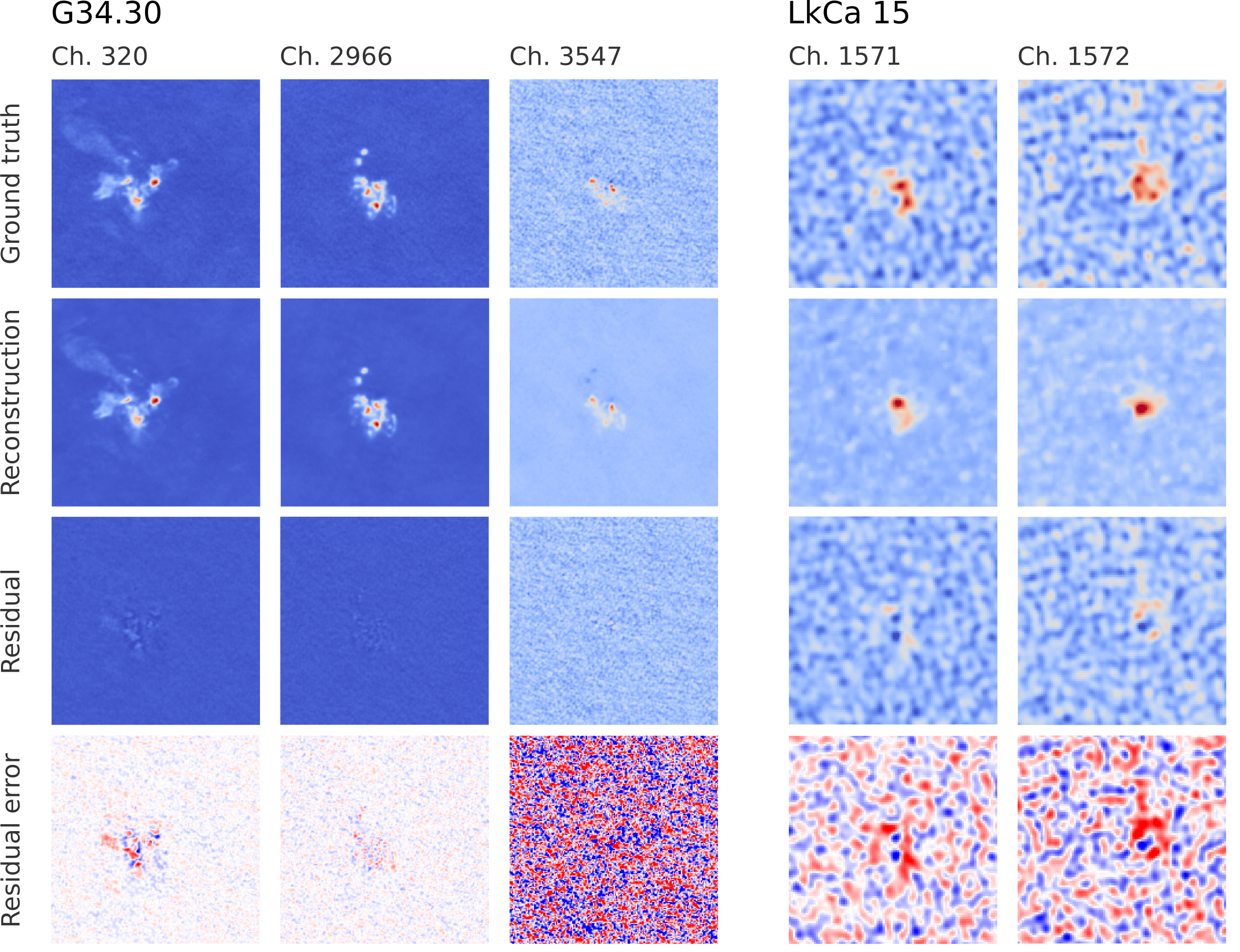}
\caption{Reconstructions, residuals, and residual errors with respect to ground truth images of select channels of G34.30 and LkCa 15 using NMF. The residual images were generated by taking the difference of the ground truth image and corresponding reconstruction. The residual errors for channels 320 and 2966 of G34.30 are 10\% of the maximum signal intensity, 40\% for channel 3547, and LkCa 15 are 30\% of the maximum signal intensity.}
\label{fig:nmf_residuals}
\end{figure*}

\begin{figure*}
\centering
\includegraphics[width=0.8\textwidth]{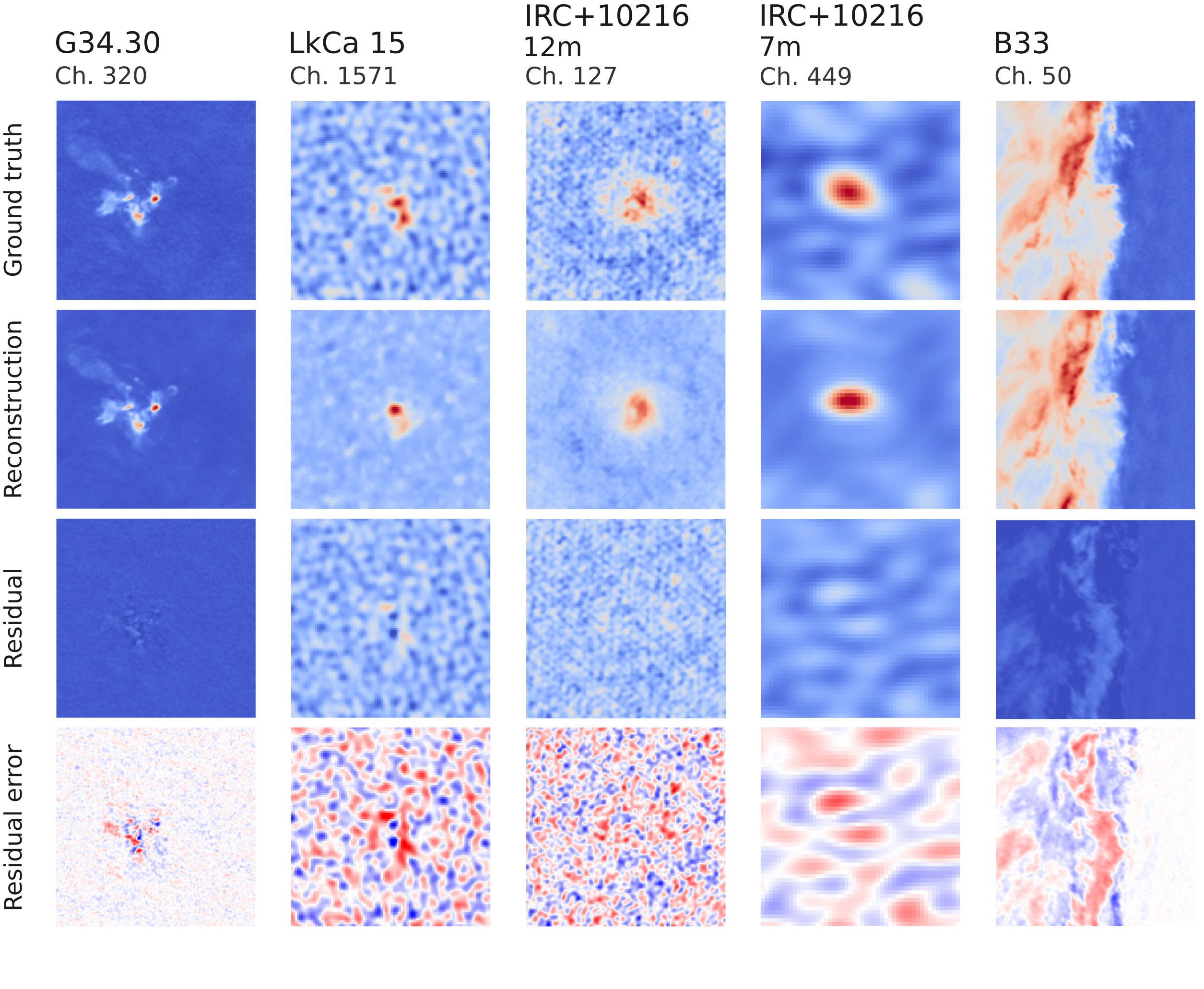}
\caption{Reconstructions, residuals, and residual errors with respect to ground truth images of select channels using NMF. The residual images were generated by taking the difference of the ground truth image and corresponding reconstruction. The residual error for channel map of G34.30 is 10\% of the maximum signal intensity, LkCa 15 30\%, IRC+10216 12m and 7m 40\%, and B33 is 20\% of the maximum signal intensity.}
\label{fig:nmf_residuals_all}
\end{figure*}

Channel map reconstructions using NMF comparing select channels of G34.30 and LkCa 15 are in Figure \ref{fig:nmf_residuals} and reconstructions for all sampled sources are provided in Figure \ref{fig:nmf_residuals_all}. No prior denoising was required, however additional considerations were made for negative values corresponding to possible absorption and baseline oversubtraction. Since NMF requires all matrix values to be positive, we addressed the presence of negative values in the dataset by splitting the initial data cube into two separate cubes: one containing positive values and the other holding the negative ones. The negative cube was transformed by taking the absolute value of its elements to simulate positive values. Both cubes were then input individually into the NMF algorithm, with the absolute value cube being multiplied by -1 to revert to its original negative state. Finally, the individual cubes were recombined to produce the final NMF output (Figure \ref{fig:block_v3}). 

A total of 50 components were used in the reconstructions for all sources except for B33 which used 5 components due to possible overfitting. Computations for all sources were performed using the local AMD workstation with execution times on the order of seconds to minutes. MSE values were comparable to those obtained with PCA with the advantage of increased computational efficiency. No prior denoising was required, however additional considerations were made for negative values corresponding to possible absorption and baseline oversubtraction. NMF was consistent with previous treatments in that the emission structure was well captured in channel maps for spectral-rich sources with high signal-to-noise ratios (i.e. G34.30). The most significant errors in the G34.30 dataset corresponded to areas of the peak emission, whereas the residual images for the B33 dataset showed minimal difference between the ground truth and lower-dimensional reconstructions, indicating potential overfitting due to the limited number of spectral channels in the image cube, however this could not be confirmed without additional cross validation testing. 

NMF demonstrated improved performance on lower-SNR and spectrally-poor datasets compared to previous methods, despite evident limitations. Similarly to Sparse PCA, NMF does not reconstruct thermal noise well, allowing for inherent denoising of low signal-to-noise spectral channels. NMF effectively captured the general emission structure in datasets with less resolved emission and greater line sparsity, yet still produced larger errors when compared to reconstructions of spectral-rich and more resolved channel maps. For instance, the emission outline for channel 1571 in LkCa 15 was reconstructed, yet the flux emission was not adequately captured throughout the emission region. This was also observed in the IRC+10216 12-m reconstruction (Figure \ref{fig:nmf_residuals_all}). 7-m IRC+10216 data was also reconstructed using NMF to sample an additional low SNR dataset. Although NMF captured resolved features and low lying emission, the orientation of the unresolved emission ``disk'' was spatially misaligned in the 7-m reconstruction. Upon inspection, this is an artifact of the noise around the disk being reconstructed as an ``average'' of the emission through out the spectral window.  

Overall, NMF accurately captured and reconstructed the most significant emission features across various astronomical datasets. While the algorithm exhibited similar limitations with lower resolution and line-poor sources, its lower computing time and ease of implementation make it a viable and accessible option for those with limited computational resources.

\begin{figure*}
\centering
\includegraphics[width=0.8\textwidth]{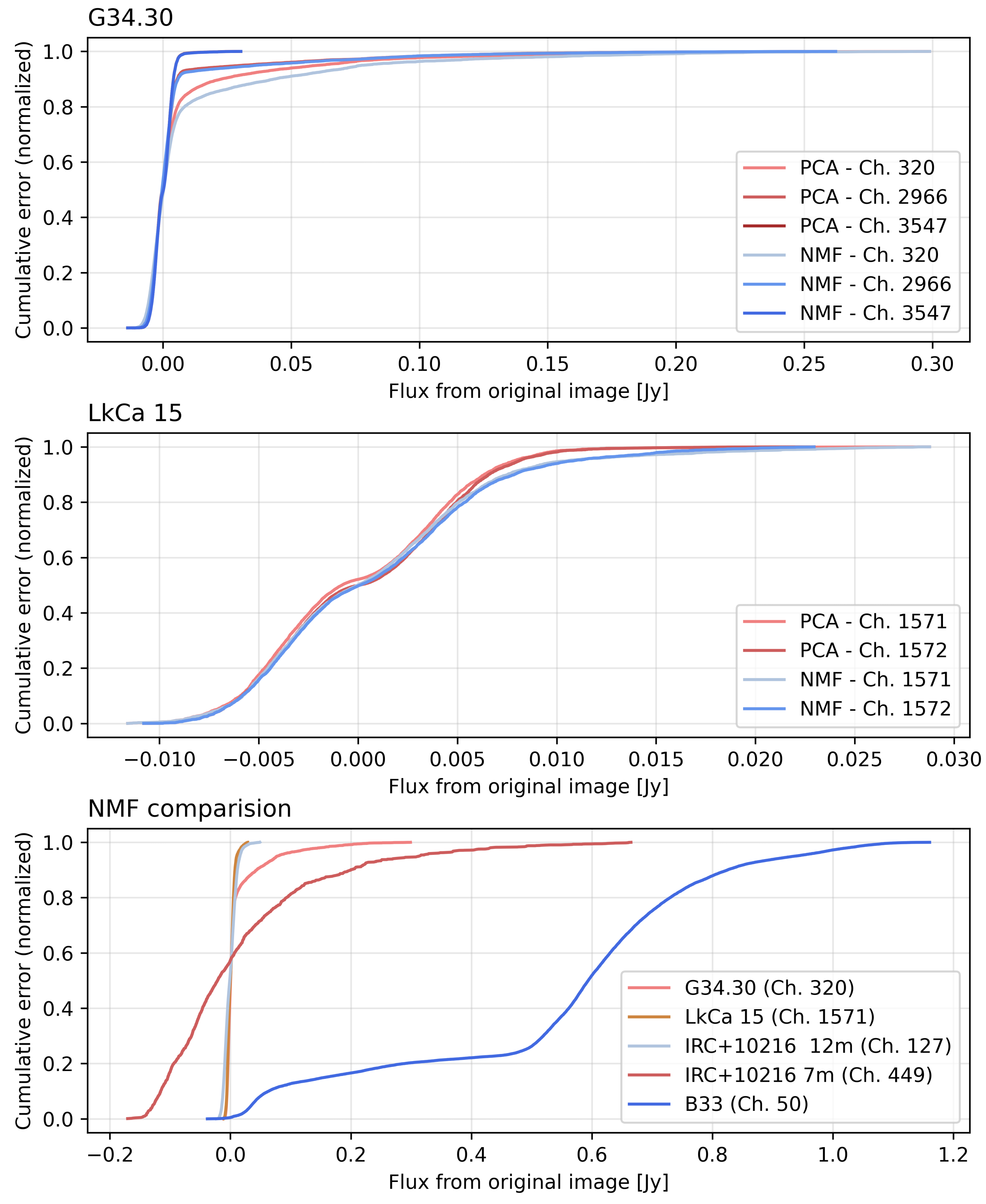}
\caption{Cumulative error in image reconstructions with respect to pixel flux of the original channel maps. The top and middle panels correspond to the errors in the G34.30 and LkCa15 reconstructions using PCA and NMF (Figures \ref{fig:pca_residuals} and \ref{fig:nmf_residuals}). The bottom panel shows the cumulative error from the select channels of each sampled source in Figure \ref{fig:nmf_residuals_all} using NMF.}
\label{fig:flux_errors}
\end{figure*}

\subsection{Autoencoder Reconstructions}

A simple convolutional autoencoder was used initially on the G34.30 and LkCa 15 data sets, then applied to the additional B33 and IRC+10216 datasets. The same architecture using 8 kernels of size 15 with a max pool factor of 2, batch size of 128, and learning rate of 0.002 were used across all datasets (as depicted in Figure \ref{fig:block_v3}). A set number of 500 epochs was also used for all datasets following an initial screening for convergence in the G34.30 data (Appendix \ref{sec:AppC}). The input and final output image sizes were consistent for all sources except for B33 as the input dimensions exceeded our computational allowance and were instead projected onto a final image size of 180x180. Computations were performed using the local NVIDIA workstation with execution times on the order of minutes for inference to hours for training, depending on the number of components used.

The reconstructions generated by the autoencoder are of higher fidelity from dataset to dataset compared to the other treatments. The corresponding MSE values are comparable to PCA and NMF in the G34.30 and LkCa 15 datasets and larger errors in the later two datasets. We found that the autoencoder performed exceptionally with datasets containing a combination of a large sample of emission instances and/or highly resolved emission. For example, the G34.30 dataset is both spectrally-dense and well-resolved, resulting in features that were well captured within latent space and reconstructed in the compressed images. The B33 dataset proved to be a specific case where the few number of channels within each spectral window were available for learning and high spectral resolution from prior group imaging resulted in near perfect reconstructions of each channel map. This may suggest overfitting by the autoencoder, however without additional cross validation training, it can only be speculated. Without direct examination of the latent space, it is difficult to gauge what information is actually contained within the embeddings, and how the information is encoded. However, the results are still viable if the end use is solely to generate compressed images. Additional reconstructions generated with fewer components to gauge the lower limits of the autoencoder compression are provided in Appendix \ref{sec:AppB}. 

The SNR of the 12-m IRC+10216 image data is lower compared to the G34.30 and B33 datasets, however the autoencoder still sufficiently reconstructed the emission, characteristic halo effect, and noise profiles. This suggests a trade-off between SNR and the number of training samples available for acceptable latent embedding. Alternatively, the limited number of channels containing emission severely limited efficient latent encoding for the LkCa 15 data. Although the dataset has high SNR, the autoencoder cannot explicitly distinguish between true emission and noise, resulting in the autoencoder trying to ``learn'' the randomized thermal noise, indicated by the lower MSE in Figure \ref{fig:mse_bar} and further suggests the need for investigations into the encoding and structure of the learned latent information.  

The number of components used also impacted the quality of the latent embeddings. When comparing the reconstructions using 50 components up to 500 components for channel 320 of G34.30, there is a noticeable ``blurring'' effect when fewer components are used, characteristic of the convolution applied by the kernel (see Figure \ref{fig:ra_residuals1} in Appendix \ref{sec:AppB}). The general shape of the emission is captured even with minimal components, but the features are not fully resolved without a larger number of components being used (n=300--500). This is indicative of the types of features that are captured in latent embeddings by the autoencoder are different compared to those of the other treatments, such as NMF.


\section{Discussion} \label{sec:disc}
\noindent {\bf Dimensionality reduction assessment by method:}

NMF is most notably used in image segmentation where strong object boundaries are essential for accurate results. In regards to its use with ALMA data, this can lead to an increased denoising of the reconstructed images due to decreased encoding of thermal noise into the latent space components---which can be desirable depending on the specific science use case. NMF proved a well rounded dimensionality reduction method with comparable results to PCA. The method demonstrated multiple benefits, including efficient computation costs (i.e. capable of running using 2 vCPUs and 12GB of RAM via Google Colab) and its straightforward implementation. NMF does not require labeled data, and significantly reduces the amount of manual preprocessing, given its relatively simple algorithm---conceptually and in practice---the requirement for all matrix values to be positive only necessitated minimal preprocessing, which was easily addressed.

Autoencoders, with their inherent flexibility, are well-suited for a wide range of applications and science use cases. The autoencoder allowed for efficient latent space encoding of important dataset features; however, autoencoders require large amounts of data for training to learn these representations, limiting their application for line sparse sources and datasets with fewer channels, such as LkCa 15 and B33, resulting in larger error in the reconstructions (Figure \ref{fig:mse_bar}). Limited sampling increases the risk of the model failing and making transfer learning inaccessible. The need for large volumes of data can cause training to be very computationally expensive. Depending on the complexity of the architecture, larger computing resources may be needed. Alternatively, in the case of B33, group imaging provided highly resolved images, leading to near perfect reconstructions. In cases where the need is for highly compressed data for storage, over-fitting is not an issue.

Of the methods sampled, PCA required the most manual dataset preprocessing, including prior denoising to limit the algorithm from generating noise back into the reconstructed images. PCA performed best on datasets with abundant spectral emission and high SNR as the method is sensitive to noise \citep{Bailey2012}. This observation naturally led to sampling Sparse PCA with the hopes that introducing sparsity to the image data would cut down on execution time while still preserving key components. The Sparse PCA reconstructions were less optimal than the traditional PCA reconstructions, leading to the possible conclusion that the intrinsic emission intensity of the image cubes may not be suitable for Sparse PCA encoding since the dynamic range of the pixel values is lower compared to typical computer vision datasets. It should be noted that a thorough gridsearch of the hyperparameters could improve the quality of the low dimensional reconstructions, however, our goal was to assess the performance of these methods in an ``off-the-shelf'' state, which maximizes generalizability across data quality and source types.
\linebreak

\begin{figure*}
\centering
\includegraphics[width=0.8\textwidth]{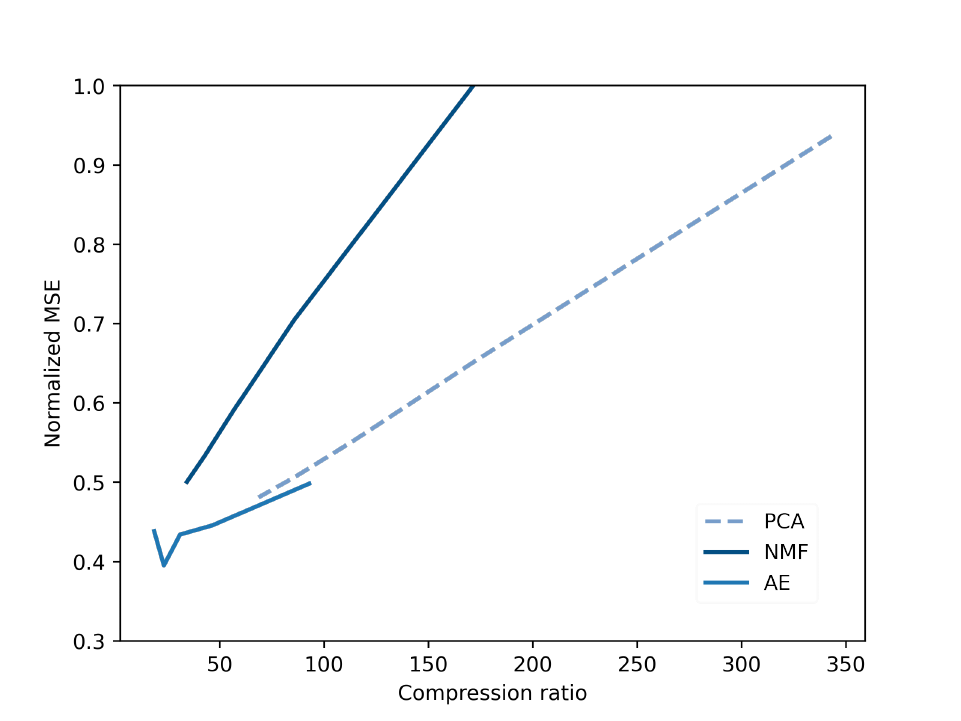}
\caption{Data compression of voxels in the G34.30 dataset using PCA, NMF, and the autoencoder. The efficacy of compression after initial downsampling was compared using the normalized MSE of reconstructions with varying number of components.}
\label{fig:compression}
\end{figure*}

\noindent {\bf Scalability of compression \& generalizability:}
 
The scalability of the sampled models is also a factor in their generalizability and ability to fit user needs. The MSE of each source provides a complementary metric to visual reconstructions in how transferable each method is for similar datasets. For example, the calculated MSEs across the sampled sources for NMF are within an order of magnitude of one another and each with well constrained error bars (Figure \ref{fig:mse_bar}). This is despite the exception of B33 where the large change in signal intensity throughout the limited number of channels resulted in larger errors, as depicted in channel 50 (Figure \ref{fig:nmf_residuals_all}) where the largest residual error corresponds to regions of highest signal flux. The autoencoder MSEs are consistently lower than the other methods, except for B33 where the limited number of channels per spectral window highlights the need of sufficient data for effective learning. Alternatively, where the errors for NMF and the autoencoder typically overlap from source to source, this is not true with LkCa 15. Despite the adequate spectral window size of the dataset, the number of channels containing true signal is low and each channel contains varying amounts and levels of noise, affecting the overall MSE as the noise and emission are captured differently between the two methods.

The extent of data compression with respect to the MSE for PCA, NMF, and the autoencoder was also considered for the G34.30 data cube following initial downsampling and emission windowing (Figure \ref{fig:compression}). Compression ratios were calculated with a varying number of components used in the reconstructions (n=10--50 for PCA and NMF, and n=100--500 for the autoencoder). The normalized MSE for all three methods increased with an increasing compression ratio with a clear linear relationship between MSE and compression ratio.

When considering the cumulative error with respect to the pixel flux of the original images, the reconstruction error of NMF and PCA is relatively higher in low signal regions (Figure \ref{fig:flux_errors}). The error decreases as the flux increases in areas corresponding to brighter emission. This is apparent for high SNR and bright, compact datasets, such as G34.30, where the majority of the reconstruction error is attributed to low signal regions and little contribution from high-flux regions where the cumulative error quickly plateaus. This behavior is also observed in low SNR datasets (i.e. LkCa 15) where the cumulative error gradually increases in comparison to high SNR datasets due to its more uniform flux distribution, but still indicates minor error contributions from high emission regions. This suggests that although the residual error is concentrated in signal-dominated regions, both NMF and PCA reconstruct bright emission well as the majority of the contributing cumulative error is from regions of low emission. This also compliments our previous assessment of NMF in its limitations of capturing thermal noise features.

Figure \ref{fig:flux_errors} also suggests there is also strong influence of the reconstruction accuracy regarding source morphology and flux intensity. The IRC+10216 and LkCa 15 datasets both depict bright, compact emission with lower SNR in their selected channels. The vast majority of the cumulative error results in these lower emission regions, namely the areas around the compact emission. This can be observed for the 7\,m IRC+10216 dataset as well, despite its significantly lower SNR and dynamical range. G34.30 itself is a hot core with compact emission, however the high SNR of the dataset allows for diffuse gas to be resolved, contributing to the previously discussed higher error within low emission and diffuse regions. B33 opposes the trends seen in the other morphologies where the cumulative error spikes at high-flux regions, which we believe further indicates possible overfitting resulting from the small number of channels in the learning space.

\section{Conclusions} \label{sec:conclusions}

This study explored the efficacy of common dimensionality reduction methods across a range of astronomical image data sizes and morphological source types. By identifying compression methods that are successful in reducing data volume and preserving important features in emission image reconstructions, we can begin to look ahead in how these treatments can be applied to different science use cases.

The results of this study demonstrated that standard dimensionality reduction methods can be applied to astronomical image cubes. We found that the majority of compression occurred prior to any algorithmic processing through simple cropping or windowing of the emission that eliminated excess pixels. ALMA datasets typically do not fill the field of view leading to unnecessary image data that compounds over the spectral window. Additionally, ALMA samples the beam with a cell size of at least five pixels per beam \citep{Hunter2023}, while sufficient Nyquist sampling is achieved with a cell size of approximately two pixels per beam. This oversampling, combined with correlation between pixels, enables additional downsampling, resulting in a total compression exceeding 90\% across the sampled ALMA datasets. This suggests that more sophisticated dimensionality reduction treatments are not always necessary for dataset compression, yet latent representations and extracted features can be useful elsewhere.

As we prepare for an increase in data load from observatory and instrument upgrades, one solution is to shift the focus to offer more tractable data products that are more readily accessible and manageable by users. Dimensionality reduction techniques such as NMF and autoencoders offer solutions for more efficient storage through compression. These methods can also be used in other tasks, such as feature extraction and image segmentation, to alleviate bottlenecks that users, especially those without access to adequate computing and cluster resources.

Upon analysis of the benchmark results and metrics, it was determined that NMF and autoencoder were the strongest methods for dimensionality reduction of ALMA image cubes of the methods sampled in this study. The success of the NMF and autoencoder models unlocks the ability to explore additional uses of the algorithms and their latent space learning such modeling non-gaussian noise profiles, auto-masking tasks, and modeling point spread functions. In comparison to the other sample methods such as NMF, PCA has slightly more disadvantages that limits its general utility as a dimensionality reduction technique with ALMA image data, most notably the additional manual preprocessing of noise and lack of interpretable components. However, it is important to note that not one specific method best serves as a ``one size fits all'' treatment across all astronomical data types and the methods chosen are dependent on each use case and end science goal.

\section*{Acknowledgements}
The National Radio Astronomy Observatory is a facility of the National Science Foundation operated under cooperative agreement by Associated Universities, Inc. Support for this work was provided to H.N.S. by the NSF through the Grote Reber Fellowship Program administered by Associated Universities, Inc./National Radio Astronomy Observatory. This paper makes use of the following ALMA data: 
\url{ADS/JAO.ALMA\#2019.1.00246.S}, 
\url{ADS/JAO.ALMA\#2015.1.00657.S}, 
\url{ADS/JAO.ALMA\#2018.1.00047.S}, 
\url{ADS/JAO.ALMA\#2019.1.00507.S} and 
\url{ADS/JAO.ALMA\#2019.1.00558.S}. 
ALMA is a partnership of ESO (representing its member states), NSF (USA) and NINS (Japan), together with NRC (Canada) and NSC and ASIAA (Taiwan), in cooperation with the Republic of Chile. The Joint ALMA Observatory is operated by ESO, AUI/NRAO and NAOJ.

%


\software{astropy \citep{astropy:2013, astropy:2018, astropy:2022},  
          Common Astronomy Software Applications \citep{Bean2022},
          Intel(R) Extension for Scikit-Learn \citep{renovate2025uxlfoundation},
          matplotlib \citep{Harris2020},
          Python \citep{hunter2007},
          PyTorch \citep{Paszke2019},
          scikit-learn \citep{Pedregosa2011}
          }



\appendix

\section{Sparse PCA} \label{sec:AppA}

Using channel 320 of the G34.30 dataset as an example, the general emission morphology was poorly reconstructed and regions of brighter emission were almost consistently oversaturated in the majority of the models. Reconstructions improved as the value of $\alpha$ was reduced and fewer components were used. Larger values of $n$ prevented the ability to properly constrain $\alpha$, leading to convergence failures and unreliable results. Additionally, larger values of $n$ resulted in oversaturation of intensities in regions of brighter emission, suggesting a trade-off in accurately capturing emission structure and intensity between $\alpha$ and $n$.

Issues with regions of lower signal-to-noise were especially pronounced in the LkCa 15 dataset, where the algorithm did not converge to a meaningful solution until $\alpha$ was significantly reduced. However, excessively lowering the value of $\alpha$ becomes trivial as this is effectively equivalent to performing standard PCA, thus negating the benefits of sparsity. 

These observations suggest Sparse PCA may not be well-suited for images dominated by noise or those with low lying emission, with the caveat that the low lying emission does not share similar morphology with more resolved emission. Inherently, this does allow for adequate ``denoising'' across the spectral window without any manual masking, suggesting sparse PCA may serve as a useful preprocessing tool when coupled with a more robust dimensionality reduction technique.   

\begin{figure*}
\centering
\includegraphics[width=1\textwidth]{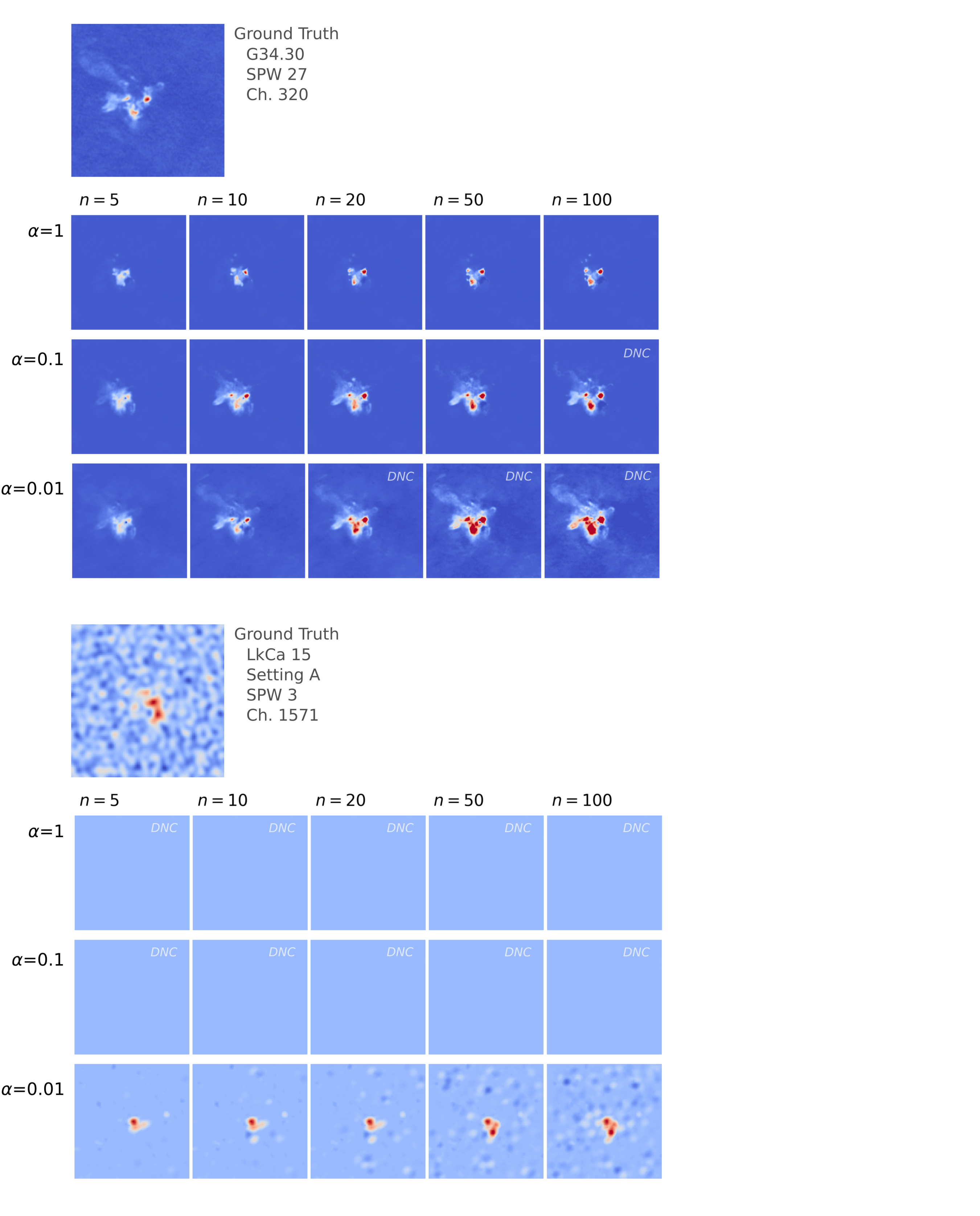}
\caption{Hyperparameter tuning of the regularization parameter $\alpha$ across varying numbers of components, $n$, for the Sparse PCA algorithm. Reconstructions shown correspond to a selected channel from the G34.30 (top) and LkCa 15 (bottom) datasets. Non-convergent results are labeled "DNC" (Did Not Converge)}
\label{fig:spca_residuals}
\end{figure*}

\newpage

\section{Autoencoder reconstructions \& residuals} \label{sec:AppB}

\begin{figure*}
\centering
\includegraphics[width=0.8\textwidth]{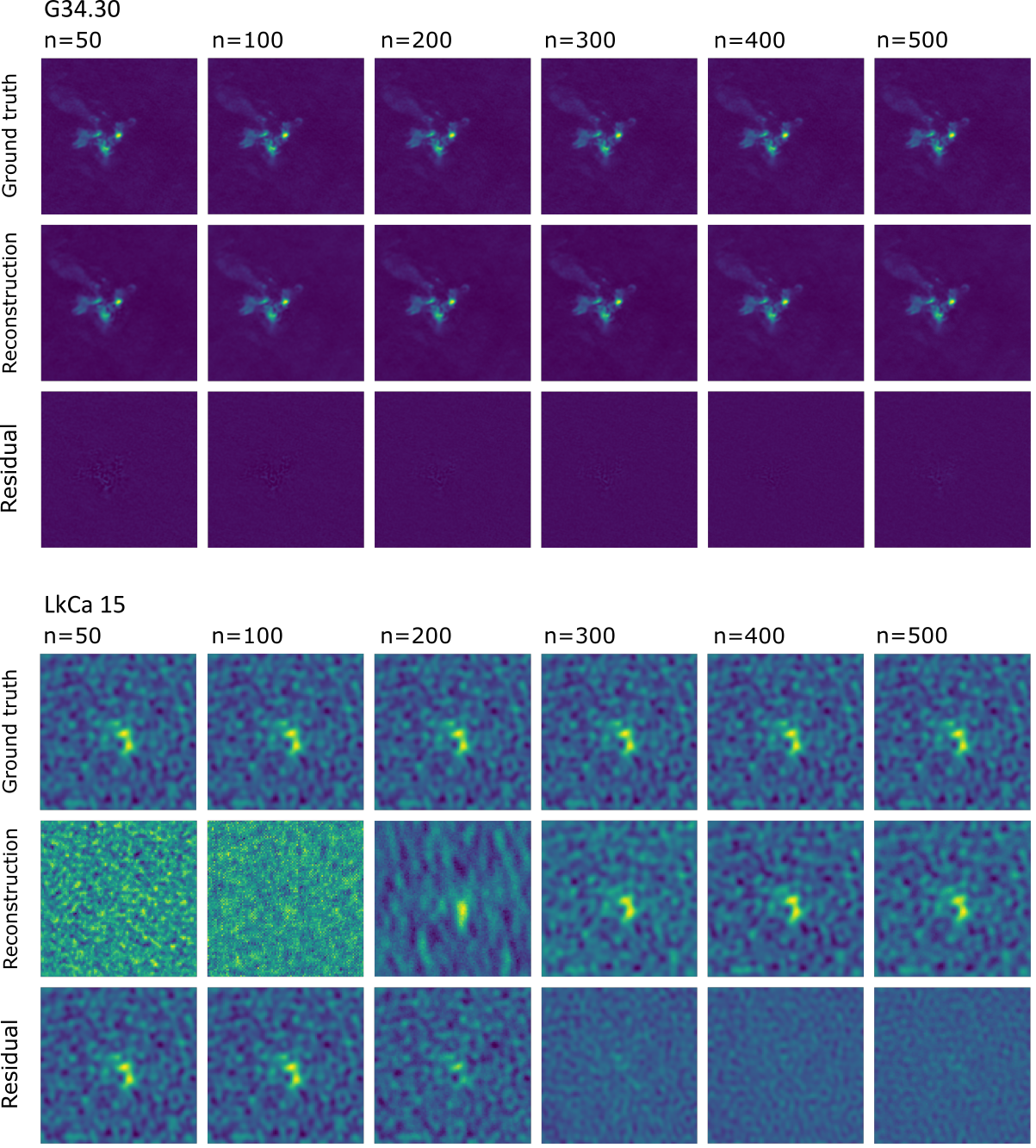}
\caption{Reconstructions, residuals, and residual errors with respect to ground truth images of select channels from the G34.30 and LkCa 15 datasets using a simple convolutional autoencoder. Reconstructions were screened with a varying number of components from $n=50$ to 500 for both datasets. The residual images were generated by taking the difference of the ground truth image and corresponding reconstruction.}
\label{fig:ra_residuals1}
\end{figure*}

\begin{figure*}
\centering
\includegraphics[width=0.8\textwidth]{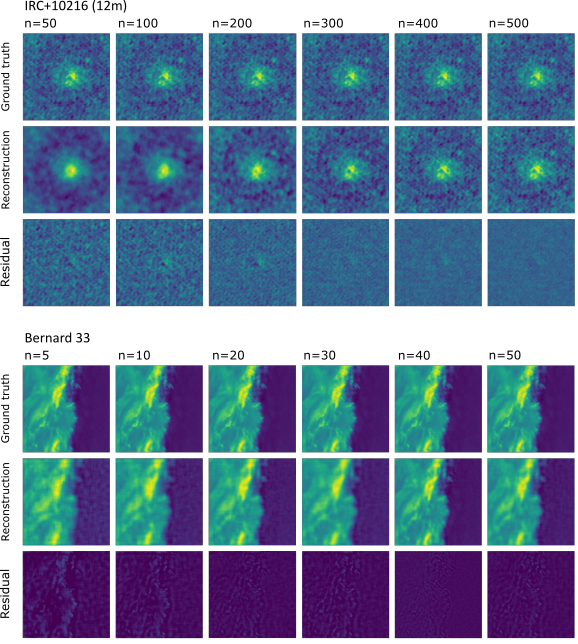}
\caption{Reconstructions, residuals, and residual errors with respect to ground truth images of select channels from the IRC+10216 and B33 datasets using a simple convolutional autoencoder. Reconstructions were screened with a varying number of components from $n=50$ to 500 for IRC+10216 and $n=5$ to 50 for B33. The residual images were generated by taking the difference of the ground truth image and corresponding reconstruction.}
\label{fig:ra_residuals2}
\end{figure*}

\section{Autoencoder convergence} \label{sec:AppC}

\begin{figure*}
\centering
\includegraphics[width=0.8\textwidth]{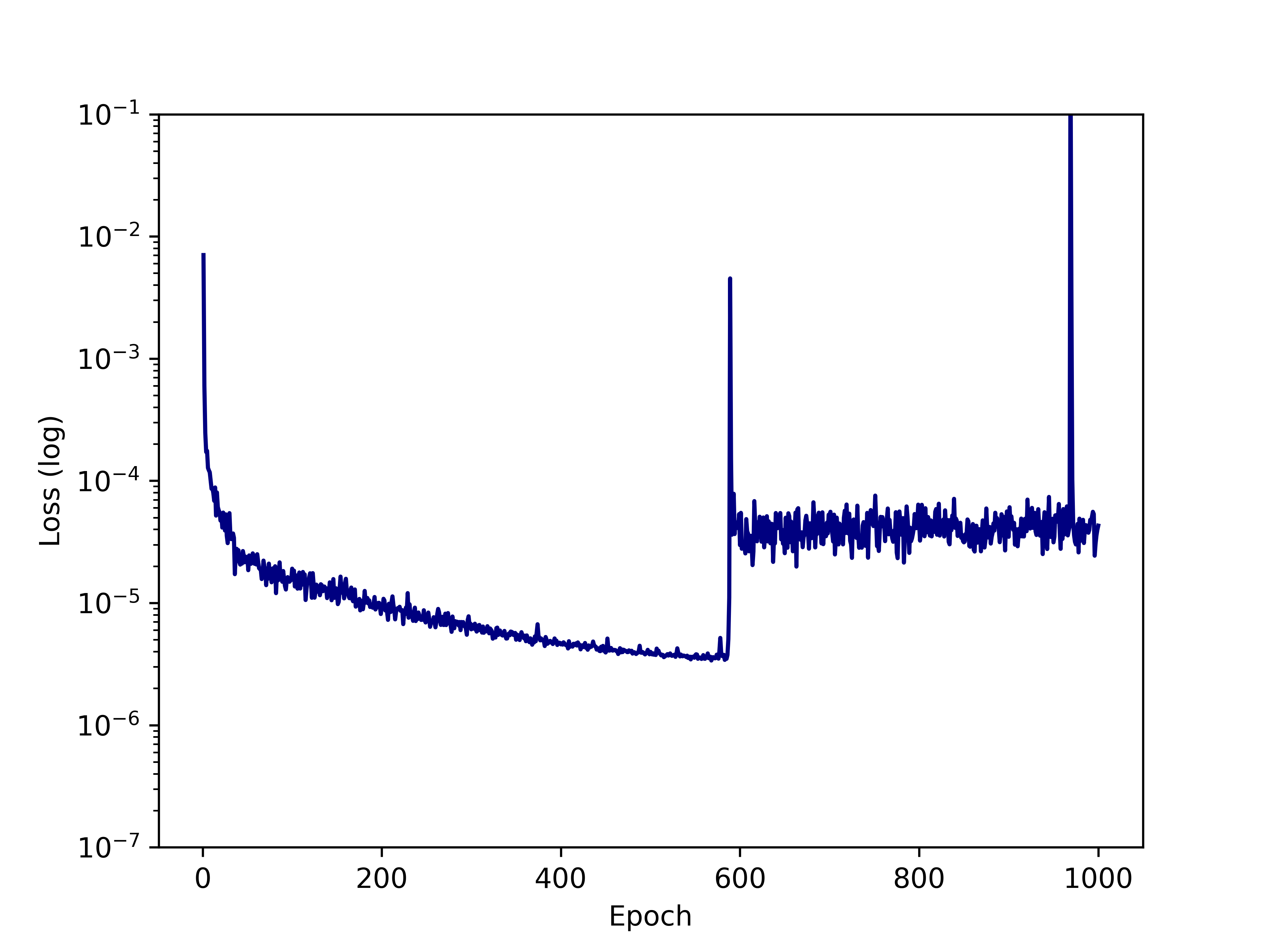}
\caption{Number of epochs with respect to loss using the G34.30 dataset.}
\label{fig:ra_converge}
\end{figure*}

A screening was performed to determine the optimal number of epochs with respect to loss to use with the radio autoencoder. Based on the results provided in Figure \ref{fig:ra_converge}, 500 epochs were used in the autoencoder reconstructions. 

\bibliography{references}{}
\bibliographystyle{aasjournal}

\end{CJK*}
\end{document}